# Privacy Risks in Mobile Dating Apps

*Full Paper*

**Jody Farnden**
University of South Australia
farja006@mymail.unisa.edu.au

**Ben Martini**
University of South Australia
ben.martini@unisa.edu.au

**Kim-Kwang Raymond Choo**
University of South Australia
raymond.choo@unisa.edu.au

## Abstract

Dating apps for mobile devices, one popular GeoSocial app category, are growing increasingly popular. These apps encourage the sharing of more personal information than conventional social media apps, including continuous location data. However, recent high profile incidents have highlighted the privacy risks inherent in using these apps. In this paper, we present a case study utilizing forensic techniques on nine popular proximity-based dating apps in order to determine the types of data that can be recovered from user devices. We recover a number of data types from these apps that raise concerns about user privacy. For example, we determine that chat messages could be recovered in at least half of the apps examined and, in some cases, the details of any users that had been discovered nearby could also be extracted.

## Introduction

Mobile devices (e.g. iOS and Android devices) and mobile applications, or apps, (e.g. healthcare apps, social networking apps, and VoIP apps) are rapidly becoming part of everyday life in both developed and developing countries. As with most new technologies, mobile devices and mobile apps can be used for criminal exploitation (Do, Martini and Choo 2015). There is a growing use of mobile devices and mobile apps to access and store sensitive and personally identifiable information (PII) data, such as healthcare and credit card details. As such, there is an increasing need for app users to have a better understanding of the privacy risks – an observation echoed in the study of Imgraben, Engelbrecht and Choo (2014).

While users access these social platforms using a wide range of mobile devices, various statistics have suggested that Google Android is the dominant platform for mobile apps (Berg Insight 2013; Research2Guidance 2013). At the time of writing, Android reportedly has an 80% market share (Lomas 2014). In the time since its release, Android has gone through much iteration and is still actively updated. For these reasons we have elected to use Android as a case study platform for our research on analysis of privacy risks in mobile apps.

Before commencement of our research, we conducted a survey of publications on the general topic of Android mobile device and mobile app user security and privacy published between 1 Jan 2009[1] and 1 May 2014. When conducting this survey, we found that there was little published work on the privacy implications of GeoSocial Networking (GSN) apps and services. A GSN app may be passive, requesting a user's location when they "check-in", such as with Facebook and Foursquare, or use a more active

---

[1]There is little research published prior to 2009 as the Android operating system (OS) was announced by the Open Handset Alliance, and made its debut to the public in the second half of 2008.

"proximity" system, where a user's location is continuously broadcast to find nearby users or locations. Proximity-based GSNs are commonly used in "friend finder" or dating networks, as these users are more concerned with finding other users near to their location, rather than just seeing users from certain locations. Meet Me for Android, a popular proximity based meet up app, boasts 145 thousand Google+ recommendations and 90 million members. Proximity based apps are not very well represented in current research trends, likely because they have only recently come to be significantly popular.

The research presented in this paper aims to contribute to an in-depth understanding of data privacy within a previously understudied app category that makes significant use of GSN, namely dating apps, both in an active data collecting capacity and profiling forensically recoverable data. Dating apps on mobile devices have come into public focus as popular dating sites, such as Plenty of Fish, are now reporting that 70% of usage takes place via a mobile phone. With the success of traditional dating apps on mobiles, many companies have taken the idea one step further and incorporated GSNs into their mobile dating app. For example, a study of the homosexual community found that many gay men surveyed had used smart phones and a GeoSocial app (Phillips et al., 2014) to facilitate casual meetings. This is not surprising when Grindr, a popular gay dating GSN, has millions of users and continues to grow daily (Grov et al., 2014).

Adding active location broadcast to these apps has raised many security concerns as users already share many sensitive details on dating profiles. With sufficiently accurate location data, it is theoretically possible to determine a user's address, track their movements and even stalk a user throughout the day (Cheung 2014). This has drawn a lot of media attention, especially to the popular Tinder app which was found to be sharing more accurate location data than intended, as users could be located to within 100 feet of their present location (Dredge 2014). This is the second reported breach of privacy in the Tinder app, the first producing the exact latitude and longitude co-ordinates of users as well as their birth dates and Facebook IDs (Seward 2013). Both security flaws were uncovered by members of the community who chose to responsibly disclose the breaches and have since been fixed, but these details could easily have been used to track users both for lawful surveillance purposes and for unlawful malicious purposed.

Aside from technical security breaches, these apps have also been used in traditional crime, such as targeted robbery and sexual assault cases (Koubaridis 2014; Wilson 2014), which have been tied to use of these services. But despite all of the hype and attention to GSNs and mobile dating apps, these services remain relatively understudied in traditional academic research, mainly being studied and analyzed only by enthusiasts. There is also limited support by professional forensic tools used by law enforcement and government agencies (e.g. EnCase, XRY, LANTERN, Paraben device seizure and ACESO), with most tools focusing on the more traditional address book and call log data. This inhibits the process of evidence collection, which is undertaken when a crime involving one or more of these apps is reported. At the time of research, we were only able to locate two prominent publications directly related to security of dating apps, both of which expose collusion exploits like the tinder example described above (Fattori et al. 2013; Qin et al. 2014). This research aims to contribute to filling this gap in contemporary research and highlighting both the privacy risks inherent in the data stored by the apps, while providing information on potential evidential artefacts for the prosecution of crimes.

## Experiment Setup

To determine the extent of artefacts generated by dating apps, we conducted a forensic analysis on nine popular proximity dating apps. We simulated user actions within the app for common usage and then took a copy of the device data. This approach was identified as the closest to a real world scenario, as any device suspected of being involved or having evidence of a crime may become subject to forensic analysis and artefacts generated by these apps may be relevant. It is important to note that any artefacts that can be extracted by the forensic process could also be extracted by a malicious attacker, which could be a potentially significant breach of user privacy. Once located, the artefacts were evaluated as to their significance.

The selected apps were chosen because they (1) allow viewing "nearby" users, and (2) ask for personal information upon sign-up or in their profiles, which includes photos, age, birthday, dating preferences, interests and often more personal details such as how many children a user has. They were also chosen for their popularity and large user bases. The final nine candidate apps were selected from the top 200 apps in the social category of the Google Play store, the top 100 apps in the Google Play store's lifestyle category and the selection of dating apps that had gained media attention as of 19th June 2014.

Of the nine apps, it was discovered that two, Badoo and Blendr, were the same app with slightly different images for the user interface, and the Blendr app requested that users follow the Badoo terms of service. As both apps used the same account and operations, it was decided that Blendr should be excluded so as not to duplicate results.

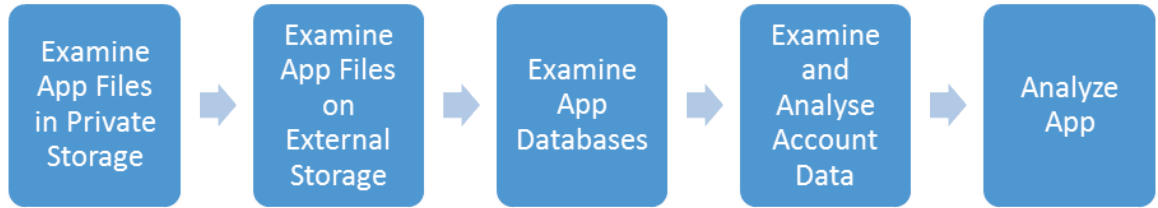

**Figure 1. App Analysis Process, adapted from Martini, Do and Choo (2015a)**

When undertaking the artefact collection we followed the process outlined in Figure 1, adapted from Martini, Do and Choo (2015a; 2015b), to ensure that we collected all relevant information saved by apps on the device. In addition to this process, as the code for these apps was mostly obfuscated and could not be directly analyzed, we captured the network traffic of each app while performing the standard actions of viewing a profile, sending and receiving a message and sending a photo through private messages if the function was available. Analysis was performed on an extraction of the data in the apps private directory using the adb pull and adb backup commands to preserve original data. Network traffic captures were recorded using a packet logger present on the device. At the end of the process, if data was found to be missing, more user actions would be performed to attempt to generate the missing data, and the process was repeated for the app in question. The experiment was performed using a Samsung Galaxy S3 GT-I9300T running Android version 4.1.2. Extracted files were analyzed using suitable tools for each file type, such as SQLite 3.8.7 to examine database files and Notepad++ for text files and unknown file types.

There are several limitations when dealing with Android in the context of examining application artefacts. The first is that many applications are essentially web wrappers for a mobile website, and do not store a great deal of data on the device. This may be circumvented by using the token directly on the website where available. The other concern when analyzing Android data forensically is the ability to access this data. This experiment maintains physical access to the device, but for this study, no passcode was set on the device, and phone data could be extracted in its entirety with root access. In real world scenarios, devices may not be this easy to access and may be encrypted. A more securely configured mobile device will generally complicate forensic efforts as forensic collection of evidential data from such devices without the cooperation of the device owner / user may be challenging (Hoog 2011), but this is a matter for cryptography specialists to explore and as such, is outside the scope of this work.

## Findings

Our findings are organized according to the procedure in Figure 1, with the exception of the "Examine App Files on External Storage" step as no data was found on external storage for the apps that we have analyzed. For reproducibility, the version of each app is listed in Table 1.

| Application Name | Version Number |
|---|---|
| Badoo | 2.46.3 (177) |
| Grindr | 2.1.1 (2072) |
| Skout | 4.3.3 (136) |
| Tinder | 3.2.1 (759) |

| Jaumo | 2.7.5 |
| --- | --- |
| Meet Me | 8.7.1 (107) |
| FullCircle | 3.0.3 (22) |
| MuiMeet | 2.7.4 (53) |

**Table 1. App Versions**

As the majority of apps analyzed use Facebook login credentials as their authentication provider, it is necessary to examine the possible use of recovered Facebook tokens. Facebook tokens obtained from these apps may be used to determine the identity of the account holder if they are using different details. By using the Facebook graphs API, the token may be submitted to obtain the users Facebook details. The graphs API may be queried with *"https://graph.facebook.com/me?access_token="* followed by the token string obtained from the app to retrieve user information. In addition, if the app has been granted access to post for the user, status updates may be posted using that apps token by sending a POST request, however many apps request the right to post for a user separately from the initial login prompt.

The following subsections describe our detailed findings for the first four apps analyzed (i.e. Badoo, Grindr, Skout and Tinder). The detailed findings for the remaining four apps have been omitted due to page constraints; however, the key findings from all eight apps are outlined in the Discussion section below.

## *Badoo*

Based on the badoo.com website database, Badoo is one of two identical apps by the Badoo Company. The second app is called "Blendr" and both website and app are identical to Badoo, both use the same membership and the privacy policy on the Blendr website currently shows the Badoo privacy policy. The Blendr app stores the exact same files as the Badoo app tested in this section. Badoo gives users the option to restrict interactions from unverified accounts. Users can verify their account via two of the following: Facebook, phone number or subscription purchase. Badoo allows sending of images in chat.

The [**private app storage path**] used by Badoo is /data/data/com.badoo.mobile.

**Examine App Files in Private Storage**

In the file "[**private app storage path**]/cache/[GUID]", we located the following data of interest:

- A list of the last messages received containing:
    - Username
    - Profile Picture URL
    - Last Message
    - Location at Suburb Level

This information can be used to link the device to an account on the service and to establish an approximate last location they communicated from. It reveals the location of the user at the time the message was sent.

In the directory "[**private app storage path**]/cache/downloader", we located the following files of interest:

- Image Files of All Viewed Profile Pictures
    - All images are "webp" or JPEG format and contain no metadata that we considered to be generally relevant.

These profile images can be used to determine who the user may have been interacting with.

In the directory "[**private app storage path**]/shared_prefs", we located the following files of interest:

- com.facebook.SharedPreferencesTokenCachingStrategy.DEFAULT_KEY.xml
- com.facebook.AuthorizationClient.WebViewAuthHandler.TOKEN_STORE_KEY.xml

Both files contain the Facebook authentication token string for the app. Facebook tokens can be used to tie account identities together and gain access to information entered into Facebook by the user.

**Examine App Databases**

The app contains one database file in the "**[private app storage path]**/databases" folder, google_analytics_v2.db, which is used to store user agreement data and was empty in our experiments.

In addition, the app contains cookies and webview databases located in "**[private app storage path]**/app_webview", which contain cookies and autofill data.

No other databases were located. The data of interest that we were able to locate in these databases is limited to the users email address.

**Analyze App**

Through network traffic monitoring we were able to collect profile pictures, chat, nearby users, user profile and device information. A preview of the last message sent or received from each user can be viewed, but not historical messages. The users profile is recoverable, along with a list of users they have declared as a "match". Device information such as device model and OS version were also noted.

## *Grindr*

Grindr is a gay dating app that shows nearby users available to chat. It allows chat between users and sending of photos. The app requires access to high GPS accuracy and will not show any user information until the GPS has reported the phone's precise location.

The [**private app storage path**] used by Grindr is /data/data/com.grindapp.android.

**Examine App Files in Private Storage**

In the folder "[**private app storage path**]/cache/Picasso-cache", we found the following files of interest:

- Image Files of All Viewed Profile Pictures
    - .0 files contain a GET request with the image URL
    - .1 files contain the image file itself

These may be used to identify possible associates the user viewed or contacted. It can be used to identify a user's associates when combined with the matching database information (described below).

In the folder "[**private app storage path**]/shared_prefs/", we located the following files of interest:

- Rules.xml which contains the following items:
    - Grindr Token
    - SessionID
    - Last Active Time (epoch value)
    - User's Email

The last active time may be used to indicate the device user's interactions, and may be teamed with a network capture to obtain location information or cross matched with the databases to obtain the last messages sent or received. The users email being present allows a link to be established with other accounts to verify identity.

**Examine App Databases**

The app contains five database files in the "**[private app storage path]**/databases" directory:

- "1dfd3ff262804a5794a90eb9d3a15b9f", an empty database with a single table named api_calls
- grindr.db which stores all app data
- webview.db for storing cookies on older Android systems
- webviewCookiesChromium.db to store cookies on Android 4.0+
- webviewCookiesChromiumPrivate, used for incognito browsing cookies

The grindr.db file is the most relevant, as it contains a significant amount of private data. The user's complete profile dataset is stored in this database, along with the profile details of all other users they have viewed. This data is located in a "profile" table, which contains privacy sensitive formation such as birth date, distance to the user on last viewing, the last time they were seen online, account information from Facebook, Instagram or Twitter if it has been provided. It contains a field for an image hash that is linked to the image files in the Picasso image cache. The Grindr database also stores a "chat" table that contains all messages sent and received by the user and the date and time they were sent.

A detailed listing of the data stored in the app databases located in the **[private app storage path]**/databases/Grindr.db directory is presented in Table 2 (see Appendix A).

**Analyze App**

Grindr sends all profile images unencrypted across the network. The user's location is sent from the device to the Grindr server with country and city data as well as the exact latitude and longitude of the user. Combined with a timeframe, this can be used to track users as long as they stay connected to the same network.

## *Skout*

Skout offers users the ability to chat with nearby users via a "shake to chat" option.

The [**private app storage path**] used by Skout is /data/data/ com.skout.android.

**Examine App Files in Private Storage**

In the "[**private app storage path**]/shared_prefs/" directory, we found the following files of interest:

- "LOGIN_PREFS.xml"
- "com.facebook.SharedPreferencesTokenCachingStrategy.DEFAULT_KEY.xml

Both files contain the Facebook token string. Facebook tokens can be used to link the app account with a Facebook account to obtain more information about the user. The token string can be used to directly request the details visible to the app.

- "LOCATION_PREFS.xml"
    - "LOCATION_LAST_SENT_TIME"

This value may be used in conjunction with the other information in databases and packet captures to map a user's location or their last interactions.

**Examine App Databases**

The app contains six database files in the "**[private app storage path]**/databases" folder.

- google_analytics_v2.db to store user agreement data
- mixpanel is an empty database

- skout.db stores all the app data
- webview.db for storing cookies on older Android systems
- webviewCookiesChromium.db to store cookies on Android 4.0+
- webviewCookiesChromiumPrivate, used for incognito browsing cookies

We determined that skout.db contains two relevant tables, skoutUsersTable and skoutMessages. The former contains a list of profiles the user has interacted with, their ID, name, profile picture URL, an ID for the last message sent and a timestamp for the last communication. SkoutMessages contains all messages sent and received.

The Skout database located in **[private app storage path]**/databases/skout.db is outlined in detail in Table 3 (see Appendix A).

**Analyze App**

Skout sends profile images of all users unencrypted over the network. It sends user data as XML files which contain username, age, birth date, interests, last seen time, location at that time with country, suburb and distance data, and a profile picture URL.

### Tinder

Tinder is a dating app that shows nearby users and allows users to mark them as a "match". Users can also take a photo and share it as a "moment" so that all of their matches can see it. Users must verify their phone number before accessing this service.

The [**private app storage path**] used by Tinder is /data/data/com.tinder.

**Examine App Files in Private Storage**

In the "[**private app storage path**]/cache/Picasso-cache" directory, we located the following files of interest:

- Image Files of All Viewed Profile Pictures
    - .0 files contain a GET request with the image URL
    - .1 files contain the image file itself

These may be used to identify possible associates the user viewed or contacted, in conjunction with database records.

In the folder "[**private app storage path**]/cache/volley", we located the following files of interest:

- JSON requests stating whether the user was matched or not
    - matchIDs from tinder.db match table
    - The date the match occurred

This establishes an association between user accounts at a certain time and can be used as a starting point that can be followed by searching app databases.

In the folder "[**private app storage path**]/shared_prefs", we located the following files of interest:

- "com.facebook.AuthorizationClient.WebViewAuthHandler.TOKEN_STORE_KEY.xml"
- "com.facebook.SharedPreferencesTokenCachingStrategy.DEFAULT_KEY.xml"

Both files contain the Facebook token string, which can be used to communicate with Facebook.

- "SP.xml"

    This file contains:

- o The user's Facebook token
- o The user's latitude and longitude
- o The user's Tinder token

Tokens may be used to gain access to the account, and to any information retrievable from Facebook by the app, as long as they have not expired. The user's latitude and longitude can be used to show where the user was when they last opened the app.

**Examine App Databases**

The app contains four database files in the "**[private app storage path]**/databases" folder:

- tinder.db stores app data
- webview.db for storing cookies on older Android systems
- webviewCookiesChromium.db to store cookies on Android 4.0+
- webviewCookiesChromiumPrivate, used for incognito browsing cookies

We determined that tinder.db contains the most data of interest, with tables for messages, analytic events, matches, photos and moments. The messages table contains all messages sent and received by the user with timestamps. The matches table lists all profiles the user connected with and the date of first contact. The moments table stores all viewable posts made by the user. The photos table contains photo ids and URLs. The analytics_Events table contains details such as the parameters sent over the network, including exact location in latitude and longitude, the type of network it is connected to and the deviceID.

The details of the Tinder database found in **[private app storage path]**/databases/tinder.db are outlined in Table 4 (see Appendix A).

**Analyze App**

All profile images viewed were present in network traffic. The user's location, sent to the Tinder server, is present in network traffic as precise latitude and longitude.

# Discussion

The information that we considered to be of significant importance, which can be retrieved from the apps, is summarized in Table 2.

| App Name | Messages | Images | Location | Email Address | Authentication Method |
|---|---|---|---|---|---|
| Badoo/Blendr | Only last received (and unencrypted) | Profile Images | Current – suburb level | No | Facebook Token |
| Grindr | Unencrypted in database | Profile Images | User exact location over network | Yes | Grindr Token |
| Skout | Unencrypted in database | HTTP URL sent with profile | Country, state, distance | No | Facebook Token |
| Tinder | Unencrypted in database | Profile Images | User exact location over network | No | Facebook Token<br>Tinder Token |

| Meet Me | Unencrypted in database | Profile and Message Images | No | No | Facebook Token |
| Jaumo | No | URLs in database | User exact location, which is sent in image filename | No | Facebook Token |
| FullCircle | Unencrypted over network (inc. images) | Over Network | Distance, over network | Yes – over network | Facebook Token |
| MiuMeet | Unencrypted over network | HTTP Links over network | User exact location, and other users in the suburb | Yes – over network | MiuMeet Token |

**Table 2. Summary of Data Retrieved**

In half of the apps presented, we were able to recover messages sent or received by the user. All apps leak profile images to some extent, but only the FullCircle app leaked images attached to messages. Grindr contained the most data about other users, while Badoo appeared to have the least personal information viewable. Messages can be used to determine if users were in contact before an event occurred, or to confirm alibis, as in all cases the messages are time stamped when they are sent and received. Images attached to these messages may be highly personal in nature and would often be embarrassing if they became publicly available. These factors should present significant privacy concerns to users of these apps. In addition, where image URLs were recoverable, the image may be stored at the URL for an undetermined period of time. Many images were available several months after the information was obtained from the extracted files, and are easily accessible by any internet browser.

Considering the personal nature of the information and images being shared over GeoSocial dating apps, it is disturbing that so much data can be so easily recovered. It is also problematic that many users are not aware how much data is being sent, stored and what their data is being used for. Many users would not appreciate their privately shared images and conversations being seen by third parties that they had not consented to. For example, section 11 of the Australian Privacy Principles outlines that entities must ensure reasonable steps are taken to protect information from misuse, loss, unauthorized access and disclosure. The recoverable data from network traffic logs and from the device's persistent storage suggest that many of these apps are not taking reasonable measures to protect this private information. Sending user's private conversations and images unencrypted, as FullCircle and MiuMeet do, is a significant breach of the user's trust.

To comply with user privacy, McIntyre and Casper (2014) suggest that consumers should be asked for their consent before any information is collected, and should have the use of the collected data made clear so they may make an informed decision about whether that data needs to be shared. Of all the apps in the experiment, only Grindr showed a user agreement on the first account login. All of the apps except for FullCircle had a privacy policy available on their respective websites, with the link to FullCircle's privacy policy encountering an error at the time of this research. In addition, many of the apps that incorporated Facebook login asked Facebook for much more data than the user is informed of during the login process, such as the videos they have shared from other services like YouTube or Vimeo and pages they have "liked". Skout goes as far as to request their friends list with birthdays, which they are not explicitly asked about, and this is concerning because the Skout user's friends are also not provided with any indication that this information is being obtained by the app.

In countries where privacy laws have a philosophy of minimal collection with a specific focus on not denying access to a service because a person does not wish to enter personal information (e.g. South Korea – see Greenleaf and Park 2014), many of these apps would breach privacy laws. For example, both Tinder and Badoo would fail this privacy requirement, as they require proof of identity in the form of Facebook login and phone number verification before a user is able to access certain features or most of the core service. The privacy laws also require organizations to keep data anonymized if possible, which is a requirement shared by the privacy acts of Germany and Australia. Using these principles as a guideline it is clear that many dating apps are collecting more information than required. In addition, South Korean privacy law states that a subject must be able to request deletion of data collected about them. Although many services offer a delete account function, these services do not provide assurances that they will remove all the information immediately or even at all. Many apps seem to shift the burden of privacy squarely onto the user, but as they do not display their privacy limitations clearly to the user, the user should not be held accountable. In a study by Liu (2014), almost all smart phone users admitted they had not sought out and read the privacy policy when using an app or website, with 95% of respondents deterred by the length of the privacy policy and 65% stating that it didn't matter whether they read it or not, as their data was still being collected. They also stated that it is very difficult to read a full privacy policy from a smart phone screen. Although Grindr displays its privacy policy upfront, it is still very difficult for the user to read, and the user is able to agree to the policy without even scrolling through the entire text.

## Conclusion

This research performed a case study analysis on nine popular Android mobile dating apps with the aim of identifying artefacts containing sensitive and personally identifiable information. These artefacts were analyzed to determine if they contained data that may be useful to malicious attackers seeking to breach user's privacy and also to forensic practitioners seeking to locate evidence of a crime. It was found that dating apps store messages or location readings on the device that can be used to reconstruct events or prove an alibi, which may aid in the prosecution of crimes involving these apps. In many cases, activities performed on the dating app could expose other members of the community, such as in the Grindr database, where there is a collection of all profiles the user has seen nearby, and the Skout app, which collected Facebook data from non-Skout users. In two cases (i.e. the FullCircle and MiuMeet apps), private images were viewable, which could be exploited by a malicious actor and this is a clear violation of user privacy.

The discovery of messages and other records, as well as Facebook tokens, are a privacy risk for users. While these experiments required physical access to the device, an attacker may be able to obtain these artefacts without physical access, which would have a greater impact on privacy (and potentially the physical security of the users, as highlighted in recent incidents involving dating apps).

In addition to the forensic implications, these privacy findings have implications for app developers, suppliers (stores) and users. App developers must consider the types of sensitive data they are collecting and storing on mobile devices that may be subject to unauthorized access (either physically or remotely) and how this data can be better protected. For example, encrypting sensitive data stored on mobile devices may not resolve the issue of unauthorized access entirely, but it at least provides another layer of difficulty for a physical attacker to break through. App suppliers should also be implementing technical procedures to detect the improper storage of sensitive data on mobile devices during the initial app validation process. Finally, the greatest responsibility falls to users to protect themselves from apps that store sensitive data without appropriate protection. Users should be cautious when selecting apps, particularly those that they will be using to store and/or transmit personal data. The Android permissions system provides some insight to the user as to the capability's apps will have to collect and transmit data (such as location and internet access privileges). However, these permissions are somewhat ineffective as they are quite coarse, for example, it would be unusual to find an app that does not request internet access, but it is difficult to know, as a user, exactly how this permission is being used (Do, Martini and Choo 2014).

Future work includes further research in aspects related to data privacy, forensics and social networks. The Facebook authentication system should be further investigated, as it is unclear whether apps are using Facebook authentication tokens correctly to avoid abuse by malicious attackers. This is not limited to attackers with physical access, as the same results should be reproducible if the token is obtained without physical access. As with other app or app category specific research, it is recommended that further research be conducted on a wider range of popular social dating apps.

# APPENDIX A – APP DATABASE LISTINGS

| "blocks" Table | |
|---|---|
| **Attribute** | **Description** |
| profile | Unique profile id |
| timeStamp | Date and time the user was blocked |
| isBlocked | Boolean to show user is blocked |
| "bodyTypeField" Table | |
| fieldID | Unique ID to link to "profile" table |
| name | Body type descriptor |
| "broadcast" Table | |
| messageID | Message id number |
| expirationDate | Date to stop displaying the message |
| "chat" Table | |
| messageID | Unique message identifier |
| Source | ProfileID of message sender |
| Target | ProfileID of message recipient |
| Timestamp | Epoch time message was sent |
| Type | Determines message type to be displayed |
| Body | Message content, can be text or image |
| Unread | Boolean determining whether to display message as new |
| Failed | Boolean to show whether message sending failed |
| "ethnicityField" Table | |
| fieldID | Unique ID to link to "profile" table |
| Name | Ethnicity |
| "flagReason" Table | |
| fieldID | Unique id number |
| Name | Reason for reporting a user |
| "imageGallery" Table | |
| messageID | ID of message image was attached to |
| mediaHash | Hash of picture file |
| Profile | Profile ID value |
| "lookingFor" Table | |
| Profile | Unique profile id |
| lookingForId | fieldID value from "LookingForField" table |
| "lookingForField" Table | |

| fieldID | Key to link from "LookingFor" table |
|---|---|
| Name | Name of category user is "looking for" (e.g., friends, chat) |
| "moderation" Table ||
| messageID | Unique message id |
| Message | Message content |
| Type | Type of moderation |
| mediaHash | Unique identifier for picture |
| Unread | Boolean value determining whether to show the message as read |
| "profile" Table ||
| profileID | Unique profile identifier |
| about | About me message entered by user |
| age | User set age |
| birthdate | User set birth date |
| isBlocked | Boolean value determining whether a user can contact you |
| isBlocker | Boolean value determining whether you can contact a user |
| bodyType | Foreign key to bodyTypeField table |
| children | how many children they have |
| displayName | The last seen display name for that profile |
| ethnicity | Foreign key to ethnicity table |
| weight | weight in grams |
| facebookID | Facebook url ending |
| headline | Users headline from profile |
| headlineDate | Last update to headline in Epoch time |
| height | Height in centimeters |
| isCurrent | value marking the logged in user |
| isFave | Boolean marking favourite users |
| Version | Version number from users app |
| profileImageHash | Hash of image (Links to image url in Picasso-cache folder) |
| relationshipStatus | Foreign key to relationshipField table |
| showAge | Boolean determining if user age is visible |
| showDistance | Boolean determining if distance to user is visible |
| twitterID | Twitter username |
| instagramID | Instagram username |

| lastSeen | Last date this person was seen in epoch time |
| --- | --- |
| profileStatus | Empty field |

**Table 3. Overview of Grindr Database Tables**

| "skoutUsersTable" Table ||
| --- | --- |
| **Attribute** | **Description** |
| userID | Unique ID |
| userName | Current username |
| picUrl | Address of image for profile |
| userLastMessageID | Last message sent to this user |
| lastMessageTimestamp | Epoch time of last message |
| "skoutMessages" Table ||
| messageID | Unique message number |
| Timestamp | Epoch time of message |
| fromUserID | Senders id |
| toUserID | Recipients id |
| chatID | ID to group chat messages together |
| Type | Type of message, normal, picture or rich text, Rich only comes from admin accounts |
| Message | The message body |
| addedFrom | Empty field |
| messageOrdered | Boolean value |

**Table 4. Overview of Skout Database Tables**

| "messages" Table ||
| --- | --- |
| **Attribute** | **Description** |
| User_id | User id of chat partner |
| Match_id | ID of match between user and chat partner |
| Client_created | Empty |
| Created | Timestamp of the message |
| Has_error | If there's an error in sending |
| Text | Body of the message |
| Viewed | If the message has been read |
| "Analytic_Events" table ||
| timestamp | Timestamp of the event |

| Name | Name of event |
|---|---|
| Params | Parameters of the event such as userID, latitude, longitude, network type and deviceId |
| "facebook_friends" Table | |
| Id | User id |
| Name | Users name |
| Avatar_url | url of profile picture |
| State | Empty |
| Tinder | Tinderid |
| "Match_requests" Table | |
| Empty table | |
| "matches" Table | |
| Id | Match id |
| User_id | User id |
| Created | Date the match was made |
| Last_activity | Date of last activity between user and matched user |
| Server_message_count | Empty |
| Touched | Whether the user has sent any messages to the match |
| Viewed | Whether the match has been viewed |
| User_name | The username of the match partner |
| Draft_msg | Empty field |
| Reported_for | Whether the user has been reported |
| Gender | 0 or 1 for male and female |
| Following | Whether the matched user is being followed or ignored |
| "Moment_likes" Table | |
| Empty Table with Date, Moment_id, Liked_by_id, Thumb_url, Has_been_viewed, Mixed_id, and By_user_id attributes | |
| "moments" table | |
| Id | Id of moment data |
| User_id | Id of user that created the moment |
| Created | Date of creation |
| Text | Any text posted with the moment |
| Photo_id | Id of image file attached |
| Filter | Filter on image file |
| Text_alignment | Any text added to image |

| Text_size | The size of the text |
|---|---|
| Text_height | The height of the text |
| Is_pending | Whether the moment has been posted publicly yet |
| Has_failed | Whether the post has failed |
| Rated_type | Whether the post is rated |
| Num_likes | The number of users who have liked the moment |
| "photos" Table ||
| Id | Id of photo |
| User_id | User id of photo owner |
| Image_url | URL of image if hosted |
| Origin_x | Image offset on x axis for cropped images |
| Origin_y | Image offset on y axis for cropped images |
| Height | Height of image |
| Width | Width of image |
| Xoffset_percent | Scaled image details |
| Yoffset_percent | Scaled image details |
| Xdistance_Percent | Scaled image details |
| Ydistance_Percent | Scaled image details |
| Photo_order | Photo index number for photo sets |
| "photo_moments" Table ||
| Id | ID of photo moment |
| Large | Url of large sized image |
| Med | Url of medium sized image |
| Orig | Url of original image |
| Small | Url of small sized image |
| thumb | Url of image thumbnail |

**Table 5. Overview of Tinder Database Tables**